\documentclass[conference]{IEEEtran}
\usepackage{xcolor}
\usepackage{graphicx}
\usepackage[utf8x]{inputenc}
\usepackage{tabularx}
\usepackage{listings}
\usepackage{verbatim}
\usepackage{fixltx2e}
\usepackage{lipsum}
\usepackage{multicol}
\usepackage{multirow}
\usepackage{soul}
\usepackage{pdfpages}

\newcommand{\Set}{\textsf{Set}}
\newcommand{\List}{\textsf{List}}
\newcommand{\Map}{\textsf{Map}}
\newcommand{\Sets}{\textsf{Sets}}
\newcommand{\Lists}{\textsf{Lists}}
\newcommand{\Maps}{\textsf{Maps}}

\usepackage{xspace}

\usepackage{algorithmic}

\usepackage{url}
\urldef{\autoseersgrant}\url{FCOMP-01-0124-FEDER-020484}
\urldef{\mygrant}\url{SFRH/BPD/73358/2010}
\urldef{\tiagosgrant}\url{SFRH/BD/30215/2006}
\urldef{\ssaappsgrant}\url{FCOMP-01-0124-FEDER-010048}
\urldef{\fatbitsgrant}\url{FCOMP-01-0124-FEDER-020532}
\urldef{\jorgesgrant}\url{BI1-2013_PTDC/EIA-CCO/120838/2010_UMINHO} 
\urldef{\ruisgrant}\url{BI2-2012 PTDC/EIA-CCO/1086613/2008}
\urldef{\christophesgrant}\url{BI3-2011 PTDC/EIA-CCO/108613/2008}
\urldef{\myhomepage}\url{http://www.di.uminho.pt/~jacome}
\urldef\joaosgrant\url{SFRH/ BPD/46987/2008}
\urldef\ssaappurl\url{http://ssaapp.di.uminho.pt}
\urldef\tlturl\url{http://code.google.com/p/2lt}
\urldef\jpaulo\url{jpaulo@{di.uminho.pt,fe.up.pt}}
\urldef\jacome\url{jacome@di.uminho.pt}
\urldef\jas\url{jas@di.uminho.pt}
\urldef\jorge\url{jorgemendes@di.uminho.pt}

\usepackage[all]{xy}
\newcommand{\arLaw}[5]{%
\xymatrix{
        #1      \ar@@/^1pc/[rr]^-{#4} &
        #5 &
        #2      \ar@@/^1pc/[ll]^-{#3}
}}

\newcommand{\arLawd}[5]{%
\xymatrix{
        #1      \ar@@/^1pc/[dd]^-{#4} \\
        #5  \\
        #2      \ar@@/^1pc/[uu]^-{#3}
}}

\newcommand{\arLawo}[5]{%
\xymatrix{
        #1 \ar@@/^1pc/[ddrr]^-{#4} & & \\ 
      & #5 & \\
    & & #2 \ar@@/^1pc/[uull]^-{#3}
}}

\begin{document}

\title{The Influence of the Java Collection Framework on Overall Energy Consumption}

\author{\IEEEauthorblockN{
		Rui Pereira\IEEEauthorrefmark{1}\IEEEauthorrefmark{2},
		Marco Couto\IEEEauthorrefmark{1}\IEEEauthorrefmark{2},
		Jácome Cunha\IEEEauthorrefmark{3},
		João Paulo Fernandes\IEEEauthorrefmark{4}, and
		João Saraiva\IEEEauthorrefmark{1}\IEEEauthorrefmark{2}
	}
	\IEEEauthorblockA{
		\IEEEauthorrefmark{1}
		HASLab/INESC TEC, Portugal
	}
	\IEEEauthorblockA{
		\IEEEauthorrefmark{2}
		Universidade do Minho, Portugal
	}
	\IEEEauthorblockA{
		\IEEEauthorrefmark{3}
		NOVA LINCS, DI, FCT, Universidade NOVA de Lisboa, Portugal
	}
	\IEEEauthorblockA{
		\IEEEauthorrefmark{4}
		RELEASE, Universidade da Beira Interior, Portugal\\
		\url{{ruipereira,marcocouto,jas}@di.uminho.pt}, \url{jacome@fct.unl.pt}, \url{jpf@di.ubi.pt}
	}
}

\maketitle

\begin{abstract}

This paper presents a detailed study of the energy consumption of the different
Java Collection Framework (JFC) implementations. For each method of an implementation in this
framework, we present its energy consumption when handling different
amounts of data. Knowing the greenest methods for each implementation,
we present an energy optimization approach for Java programs: based on
calls to JFC methods in the source code of a program, we select the
greenest implementation. Finally, we present preliminary results of
optimizing a set of Java programs where we obtained $6.2\%$ energy
savings.

\end{abstract}
\IEEEpeerreviewmaketitle

\section{Introduction}

The increasing energy costs related to ICT in organizations~\cite{harmon2009sustainable}, and society's environmental concerns, 
are changing the way both computer manufacturers and software engineers
develop their products. While in the previous century, improving execution time
was the main goal when developing hardware/software, and thus
programming languages and their compilers were designed to produce
fast systems, nowadays energy consumption is becoming the bottleneck
of such systems. As a consequence, powerful libraries offered by
programming languages and their compiler optimizations have to
consider this new reality.

In this paper we conduct a detailed study in terms of energy consumption
of the widely used \textit{Java Collections Framework} (JCF) library~\footnote{\url{https://docs.oracle.com/javase/7/docs/technotes/guides/collections/index.html}}. We consider
three different groups of data structures, namely Sets, Lists and Maps, and for each
of these groups, we study the energy consumption of each of its different implementations
and methods. We exercise and monitor the energy consumed by each of
the API methods when handling low, medium and big data sets. 

A first
result of our study is a quantification of the energy spent by each method of each implementation, for each of the data structures we consider. This energy-awareness can not
only be used to steer software developers in writing greener
software, but also in optimizing legacy code. In fact, we have
used/validated this quantification by semi-automatically optimizing the energy
consumption of a set of similar software systems. 

As a second result, we statically
compute which implementations and methods are used in the source code of such projects,
and then look up the energy consumption data to find which
equivalent implementation has the lowest energy consumption for those
specific methods. Finally, we manually transform the source code to
use the ``greenest'' implementation.  Our preliminary results show that
energy consumption decreased in all the optimized software systems that we tested,
with an average energy saving of $6.2\%$.

With our work we are answering the following research questions:

\begin{itemize}

\item
\textbf{RQ 1} - Can we define an energy consumption quantification of Java data structures and
their methods?

\item
\textbf{RQ 2} - Can we use such quantification to decrease the energy
consumption of software systems?

\end{itemize}

This paper is organized as follows: Section~\ref{sec:rankingJCF}
contains our analysis of the energy consumption of the different Java
Collection Framework implementations. In Section~\ref{sec:Optimizing}
we describe our methodology to optimize Java programs and its
application to five Java programs. Section~\ref{sec:Threats} presents
the validity threats for our analysis. Next, we present related and
future work (Sections~\ref{sec:Related} and \ref{sec:FutureWork},
respectively), and in Section~\ref{sec:Conclusions} we present the
conclusions of our work.

\section{Towards a Ranking of Java Implementation's Methods}
\label{sec:rankingJCF}
One of our goals is to compare the energy consumption of different Java implementations of the same abstract data structures.
To do this, we designed an experiment that simulates different kinds
of uses of such structures. 
In this section we present the design, execution, and results of that simulation. 


\subsection{Design}

Our experiment design is inspired by the one used in~\cite{manotas2014seeds}, since our analysis also considers a simple scenario of storing, retrieving, and deleting String values 
in the various collections.

\paragraph{JCF Data structures}
The most classical way to separate Java data structures is into groups which implement the interfaces \Set\footnote{\url{https://docs.oracle.com/javase/7/docs/api/java/util/Set.html}}, \List\footnote{\url{https://docs.oracle.com/javase/7/docs/api/java/util/List.html}}, or \Map\footnote{\url{https://docs.oracle.com/javase/7/docs/api/java/util/Map.html}}, respectively.
 
This separation indeed makes sense as each interface has its own distinct properties and purposes (for example, there is no ordering notion under \Sets). 
 
In our study, a few implementations were not evaluated as they are quite particular in their usage
and could not be populated with strings.
In particular, \textit{JobStateReasons} (\Set) only accepts \textit{JobStateReason} objects,
\textit{IdentityHashMap} (\Map) accepts strings but compares its elements with the identity function, 
and not with the \textit{equals} method.


Given these considerations, we evaluated the following implementations:

\begin{description}
\item[\Sets] \textit{ConcurrentSkipListSet, CopyOnWriteArraySet, HashSet, LinkedHashSet, TreeSet}
\item[\Lists] \textit{ArrayList, AttributeList, CopyOnWriteArrayList, LinkedList, RoleList, RoleUnresolvedList, Stack,  Vector}
\item[\Maps] \textit{ConcurrentHashMap, ConcurrentSkipListMap, HashMap, Hashtable, IdentityHashMap, LinkedHashMap, Properties, SimpleBindings, TreeMap, UIDefaults, WeakHashMap}
\end{description}

\paragraph{Methods}

To choose the methods to measure for each abstraction, we looked at the generic API list for the corresponding interface.


From this list, we chose the methods which performed insertion, removal, or searching operations on the data structures, along with a method to iterate and consult all the values in the structure. In some methods (e.g. \textsf{containsAll} or \textsf{addAll}), a second data structure  is needed.

\begin{description}
\item[\Sets] \textit{add, addAll, clear, contains, containsAll, iterateAll, iterator, remove, removeAll, retainAll, toArray}
\item[\Lists] \textit{add, addAll, add (at an index), addAll (at an index), clear, contains, containsAll, get, indexOf, iterator, lastIndexOf, listIterator, listIterator (at an index), remove, removeAll, remove (at an index), retainAll, set, sublist, and toArray}
\item[\Maps] \textit{clear, containsKey, containsValue, entrySet, get, iterateAll, keySet, put, putAll, remove, and values}
\end{description}

\paragraph{Benchmark}

To evaluate the different implementations on each of the described methods, we started by creating and populating objects with different sizes for each implementation.~\footnote{We will refer to the population size of an object as \textit{popsize}.}

We considered initial objects with 25.000, 250.000, and 1.000.000 elements, providing our analysis with multiple orders of magnitude of measurement. This will allow us to better 
understand how the energy consumption scales in regards 
to population size. 




When a second data structure is required, we have adopted for it a size\footnote{We will refer to the size of each such structure as \textit{secondaryCol}.} of 10\% the \textit{popsize} of the tested structure, containing half existing values from the tested structure and half new values, all shuffled.


Table~\ref{setMethods} briefly summarizes how each method is tested for the \Set collection. The tests for the other collections are similar, and their full description can be found in this paper's appendix.

\begin{table}[!htb]
\centering
\caption{Test description of Set methods}
\label{setMethods}
\begin{tabular}{ll}
\hline
Method      & Description of the test for the method          \\ \hline
add         & add popsize/10 elements. half existing, half new      \\
addAll      & addAll of secondaryCol 5 times                        \\
clear       & clear 5 times                                         \\
contains    & contains popsize/10 elements. half existing, half new \\
containsAll & containsAll of secondaryCol 5 times                   \\
iterateAll  & iterate and consult popsize values                    \\
iterator    & iterator popsize times                                \\
remove      & remove popsize/10 elements. half existing, half new   \\
removeAll   & removeAll of secondaryCol 5 times                     \\
retainAll   & retainAll of secondaryCol 5 times                     \\
toArray     & toArray 5 times                                       \\ \hline
\end{tabular}
\end{table}

\subsection{Execution}\label{sec:execution}
To analyze the energy consumption, we first implemented our data structure analysis design as an energy benchmark framework. This is one of our contributions, and can be found at \url{https://github.com/greensoftwarelab/Collections-Energy-Benchmark}. This implementation is based on a publicly available micro-benchmark\footnote{https://dzone.com/articles/java-collection-performance} which evaluates the runtime performance of different implementations of the Collections API, and has been used in a previous study to obtain energy measurements~\cite{manotas2014seeds}.

To allow us to record precise energy consumption measurements from the CPU, we used Intel's Runtime Average Power Limit (RAPL)~\cite{david2010rapl}. RAPL is an interface which allows access to energy and power readings via a model-specific register. Its precision and reliability has been extensively studied~\cite{Hahnel2012,DBLP:journals/micro/RotemNAWR12}. More specifically, we used jRAPL~\cite{liu2015data} which is a framework for profiling Java programs using RAPL. Using these tools permitted us to obtain energy measurements on a method level, allowing us a fine grained measurement.

We ran this study on a server with the following specifications: Linux 3.13.0-74-generic operating system, 8GB of RAM, and Intel(R) Core(TM) i3-3240 CPU @ 3.40GHz. This system has no other software installed or running other than necessary to run this study, and the operating system daemons. Both the Java compiler and interpreter were versions 1.8.0\_66. 

Prior to executing a test, we ran an initial ``warm-up'' where we instantiated, populated (with the designated \textit{popsize}), and performed simple actions on the data structures. Each test was executed 10 times, and the average values for both the time and energetic consumption were extracted (of the specific test, and not the initial ``warm-up'' as to only measure the tested methods) after removing the lowest and highest 20\% as to limit outliers.

\subsection{Results}

\begin{figure*}[h!]
\centering
\begin{minipage}[b]{.49\textwidth}
\centering
\includegraphics[width=0.8\columnwidth]{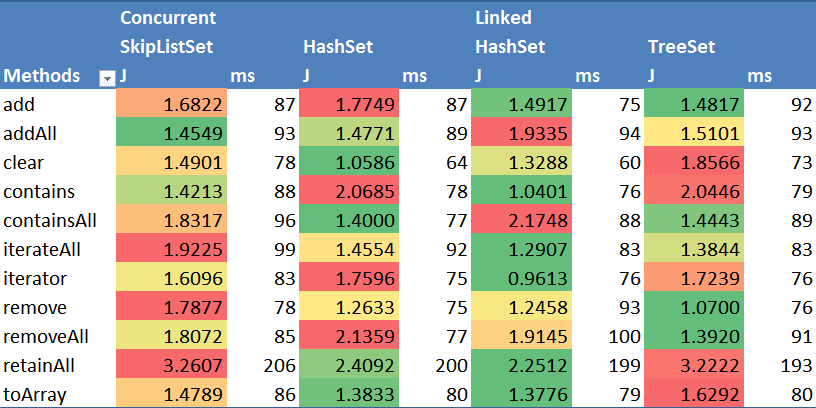}
\caption{Set results for population of 25k}	
\label{fig:set25k}
\end{minipage}\hfill
\begin{minipage}[b]{.49\textwidth}
\centering
\includegraphics[width=0.8\columnwidth]{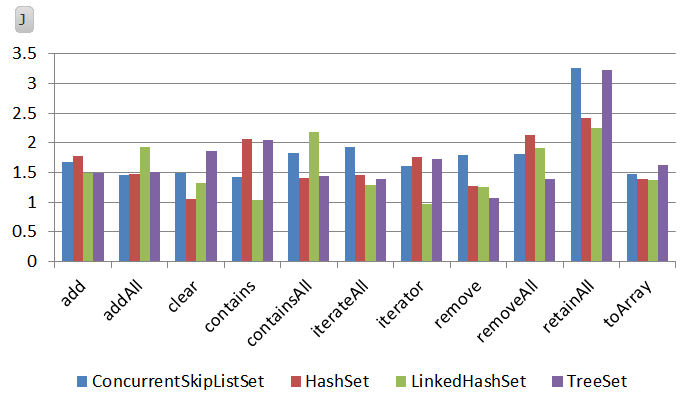}
\caption{Set results graph for population of 25k}
\label{fig:setChart}
\end{minipage}
\end{figure*}

\begin{figure*}[h!]
	\begin{center}
		\includegraphics[width=0.8\textwidth]{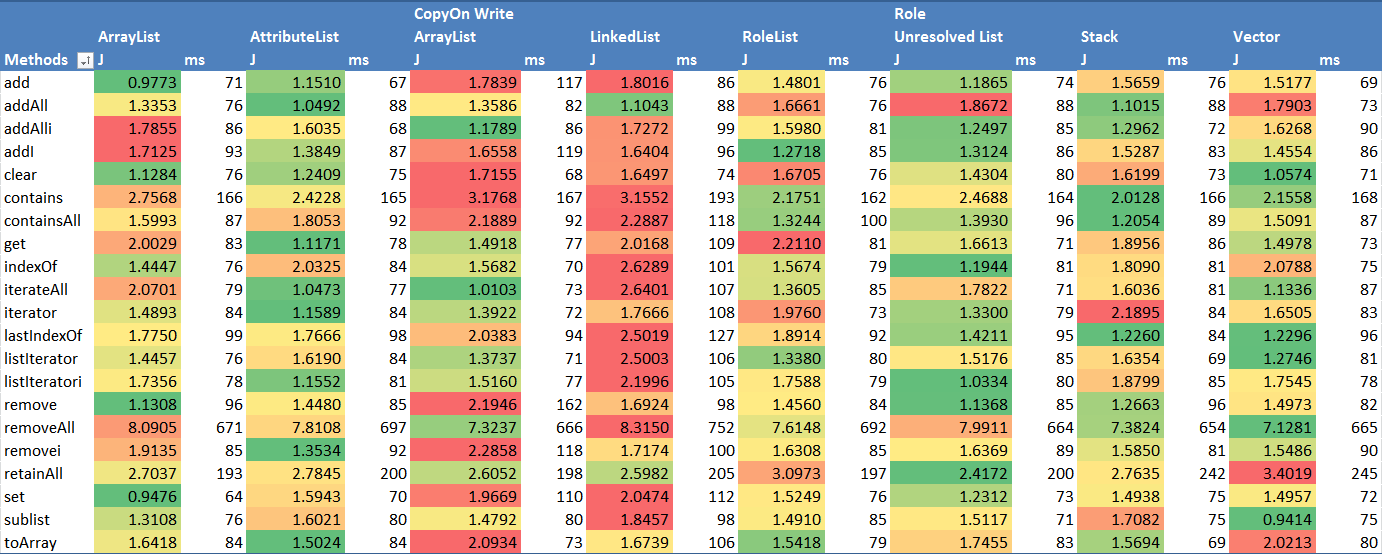}
	\end{center}
\caption{List results for population of 25k}	
\label{fig:list25k}
\end{figure*}

\begin{figure*}[h!]
	\begin{center}
  \includegraphics[width=0.9\textwidth]{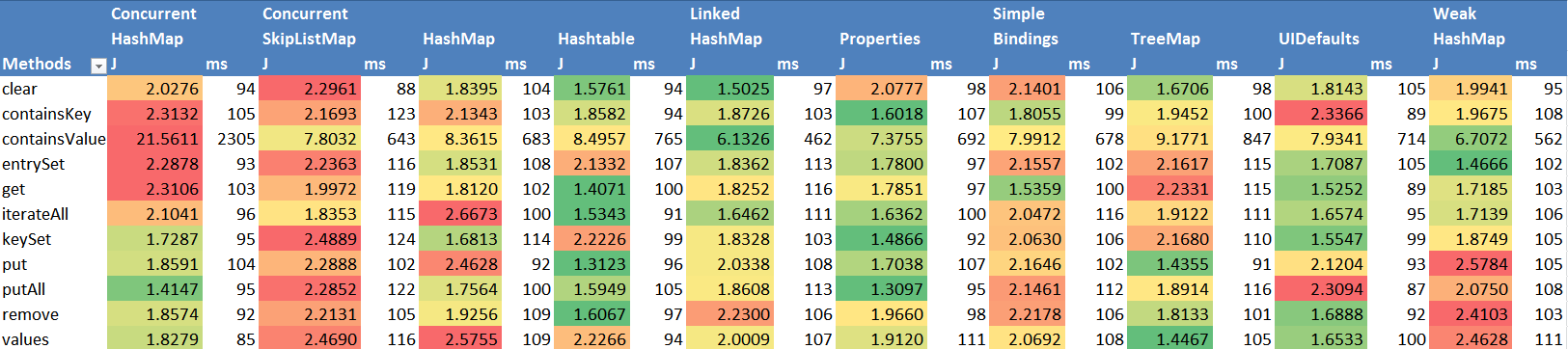}
  	\end{center}
  \caption{Map results for population of 25k}
  \label{fig:map25k}	
\end{figure*}

This section presents the results we gathered from the experiment. We highly recommend and assume the images are being viewed in color. Figures~\ref{fig:set25k}, \ref{fig:list25k}, and \ref{fig:map25k} represent the data for our analyzed \Sets, \Lists, and \Maps ~respectively. Each row in the tables represents the measured methods, and for each analyzed implementation, we have two columns representing the consumption in Joules(\textit{J}) and execution time in milliseconds(\textit{ms}). Each row has a color highlight (under the \textit{J} columns) varying between a Red to Yellow to Green. The most energetically inefficient implementation for that row's method (the one with the highest consumed Joules) is highlighted Red. The implementation with the lowest consumed Joules (most energetically efficient) is highlighted Green. The rest are highlighted depending on their consumption values when compared to the most inefficient and efficient implementation, and colored accordingly in the color scale.

\begin{figure*}[h!]
	\begin{center}
		\includegraphics[width=\textwidth]{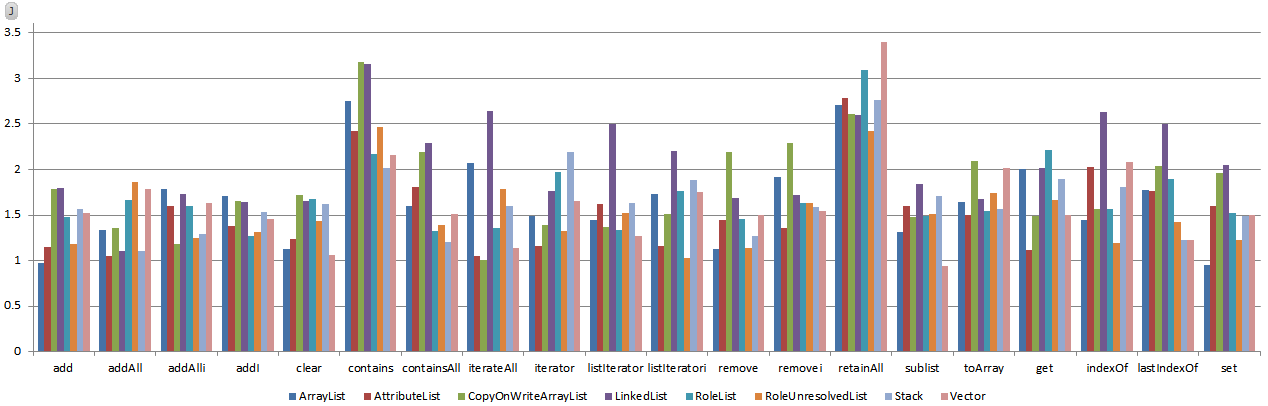}
	\end{center}
\caption{List results graph for population of 25k without removeAll}	
\label{fig:listChart}
\end{figure*}

Figures~\ref{fig:setChart}, and \ref{fig:listChart} are a graphical representation of the data for our analyzed \Sets, and \Lists. The Y-Axis represents the consumption in Joules, and the X-Axis represents the various measured methods. Each column represents a specific analyzed implementation.

The \textit{CopyOnWriteArraySet} implementation was discarded during the experiment execution as the tests did not finish in a reasonable amount of time. We also omitted the \textit{removeAll} method data from Figure~\ref{fig:listChart} as to visually improve the readability of the graph. For the full representation of the data/graphs of the other two population sizes, please consult the appendix. From our data, we can draw interesting observations:

\begin{itemize}

\item Looking at the \Set results for population of 25k data (shown in Fig~\ref{fig:set25k})
we can see that \textit{LinkedHashSet} includes most of the energetically efficient methods. Nevertheless, one can easily notice that it is also the most inefficient with the \textit{addAll} and \textit{containsAll} methods.

\item Figure~\ref{fig:list25k} presents the \List results for population of 25k. Both \textit{RoleUnresolvedList} and \textit{AttributeList} contain the most efficient methods. Interesting to point out that both of these extend \textit{ArrayList}, which contains less efficient methods, and very different consumption values in comparison with these two. We can also clearly see that \textit{LinkedList} is by far the most inefficient \List implementation.

\item In Figure~\ref{fig:map25k}, we can see that \textit{Hashtable}, \textit{LinkedHashMap}, and \textit{Properties} contain the most efficient methods, and with no red methods. Interesting to note is that while the Properties data structure is generally used to store project configuration data or settings, it produced the very good results for our scenario of storing string values.

\item The concurrent data structure implementations (\textit{ConcurrentSkipListSet, CopyOnWriteArrayList, ConcurrentHashMap, ConcurrentSkipListMap,} and the removed \textit{CopyOnWriteArraySet}) perform very poorly. As such, these should probably be avoided if a requirement is a low consuming application.

\item  One can see cases where a decrease in execution time translates into a decrease in the energy consumed as suggested by~\cite{yuki2014folklore}. For instance in Figure~\ref{fig:map25k}, when comparing \textit{Hashtable} and \textit{TreeSet} for the \textsf{get} method, we see that \textit{Hashtable} has both a lower execution time and energy consumption. As observed by~\cite{trefethen2013energy,pinto2014understanding}, cases where an increase in execution time brings about a decrease in the energy consumed can also be seen, for example in Figure~\ref{fig:map25k} when comparing \textit{HashMap} and \textit{Hashtable} for the \textsf{keySet} method. As such, we cannot draw any conclusion of the correlation between execution time and energy being consumed.

\item Different conclusions can be drawn for the 250k and 1m population sizes (which can be seen in the appendix). This also shows that the energy consumption of different data structure implementations scale differently in regards to size. What may be the most efficient implementation for one population size, may not be the best for another.

\end{itemize}

\section{Optimizing Energy Consumption of Java Programs}
\label{sec:Optimizing}
The results presented in the previous section may allow software
developers to develop more energy efficient software. In this section we present a
methodology to optimize, at compilation time, existing Java
programs. This methodology consists of the following steps:

\begin{enumerate}
\item \textit{Computing which implementation/methods are used in the programs}

\item \textit{Looking up the appropriate energy tables for the used implementation/methods}

\item \textit{Choosing the most efficient implementation based on total energy}

\end{enumerate}

In the next subsection, we describe in detail how we applied this
approach and how it was used to optimize a set of equivalent Java
programs.

\subsection{Data Acquisition}

First, we obtained several Java projects from an object-oriented course
for undergraduate computer science students. For this course, students
were asked to build a journalism support platform, where users
(Collaborators, Journalists, Readers, and Editors) can write articles
(chronicles and reports), and give likes and comments. Along with
these different platform implementations, we obtained seven test cases
which simulated using the system (registering, logging in, writing
articles, commenting, etc.). The size of users, articles, and comments
varied between approximately 2.000 and 10.000 each for each different
test case and each entity. These projects had an average of 36 classes, 104 methods, and 2.000 lines of code.

Next we discuss the optimization of five of those projects, where we
semi-automatically detected the use of any JCF implementation (both efficient and inefficient implementations), and which were the used methods for each implementation. 

\subsection{Choosing an energy efficient alternative}

To try to optimize these projects based on the data structures and their used methods, we looked at our data for the 25k population. We chose this one, as it is the one which is closest to the population used in the test cases (which was between 2.000 and 10.000 for each different entity).

For each detected data structure implementation, we selected the used methods, and chose our optimized data structure based on the implementation which consumed the least amount of energy for this specific case. 

\begin{figure*}[h!]
	\begin{center}
		\includegraphics[width=0.65\textwidth]{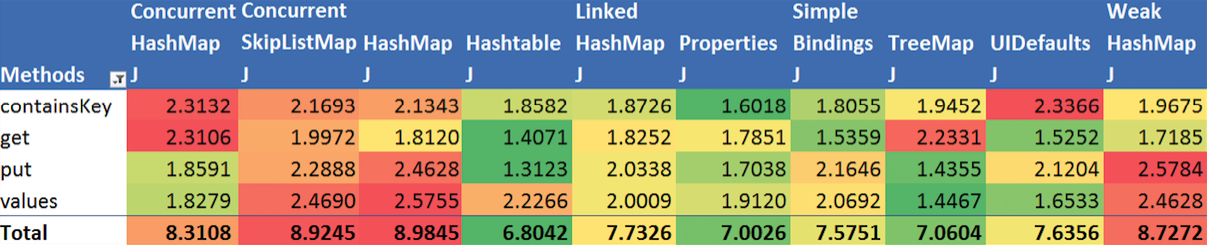}
	\end{center}
\caption{Choosing optimized Map for Project 1}	
\label{fig:optMap}
\end{figure*}

\begin{figure*}[h!]
	\begin{center}
		\includegraphics[width=0.55\textwidth]{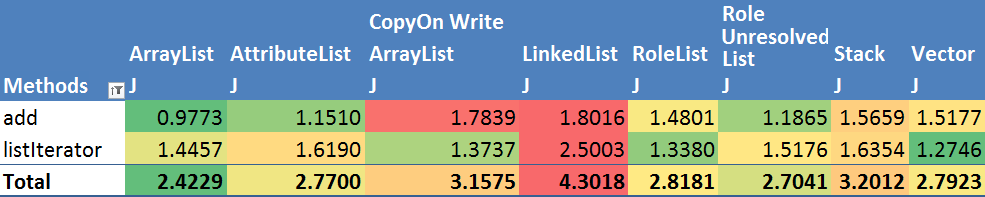}
	\end{center}
\caption{Choosing optimized List for Project 1}	
\label{fig:optList}
\end{figure*}

Figures~\ref{fig:optMap} and~\ref{fig:optList} show the data used to make our decision for Project 1, where \textit{Hashtable} was used in place of \textit{TreeMap} (as \textit{Hashtable} was the most efficient implementation in this scenario with 6.8J), and \textit{ArrayList} was used in place of \textit{LinkedList} (as \textit{ArrayList} was the most efficient implementation with 2.4J). 

Table~\ref{tab:optimization} details the 5 Projects, their originally used data structure implementations, new implementation, and used methods for the implementations.

\begin{table}[h!]
\centering
\caption{Original and optimized data structures, and used methods for each project}
\label{tab:optimization}
\begin{tabular}{c|cc|c}
\multicolumn{1}{l|}{} & \multicolumn{2}{c|}{\textbf{Data Structures}} & \multicolumn{1}{l}{} \\
\textbf{Projects} & \textbf{Original} & \textbf{Optimized} & \textbf{Methods} \\ \hline
\multirow{2}{*}{1} & TreeMap & Hashtable & \{containsKey, get, put, values\} \\
 & LinkedList & ArrayList & \{add, listIterator\} \\ \hline
2 & HashMap & Hashtable & \{containsKey, get, put, values\} \\ \hline
3 & LinkedList & ArrayList & \{add, addAll, iterator, listIterator, remove\} \\ \hline
\multirow{2}{*}{4} & LinkedList & AttributeList & \{add (at an index), iterator\} \\
 & HashMap & Hashtable & \{containsKey, get, put\} \\ \hline
\multirow{2}{*}{5} & HashMap & Hashtable & \{containsKey, get, put\} \\
 & TreeSet & LinkedHastSet & \{add, iterator\}
\end{tabular}
\end{table}

\subsection{Pre-energy measurement setup}
Now that we have chosen our energy efficient alternative, we need to change the projects to reflect this. The source code was manually altered to use the chosen implementations. Finally, we verified that the program maintained the original consistency and state by verifying if the outputs and operations produced by these two versions did not change.

\subsection{Energy measurements}

To measure the original, and optimized projects, we followed the same methodology detailed in Section:~\ref{sec:execution} \textit{Execution}. We executed the seven test cases in the same server, and using jRAPL obtained the energy consumption measurements. Each test was also executed 10 times, and the average values (after removing the 20\% highest and lowest values) were calculated.

\subsection{Results}

Table~\ref{tab:optimizationResults} presents, for each project, the energy consumption in Joules (J), and execution time in milliseconds (ms) for both the original and optimized implementations. The last column shows the improvement gained after having performed the optimized implementations for both the consumption and execution time.

\begin{table}[h!]
\centering
\caption{Results of pre and post optimization}
\label{tab:optimizationResults}
\begin{tabular}{c|llll|ll}
\multicolumn{1}{l|}{\textbf{}} & \multicolumn{4}{c|}{\textbf{Data Structures}} & \textbf{} &  \\
\textbf{Projects} & \multicolumn{2}{c}{\textbf{Original}} & \multicolumn{2}{c|}{\textbf{Optimized}} & \multicolumn{2}{c}{\textbf{Improvement}} \\ \hline
\multicolumn{1}{l|}{} & \multicolumn{1}{c}{J} & \multicolumn{1}{c|}{ms} & \multicolumn{1}{c}{J} & \multicolumn{1}{c|}{ms} & \multicolumn{1}{c}{J} & \multicolumn{1}{c}{ms} \\ \cline{2-7} 
1 & 23.744583 & \multicolumn{1}{l|}{1549} & 22.7071302 & 1523 & 4.37\% & 1.68\% \\
2 & 24.6787895 & \multicolumn{1}{l|}{1823} & 23.525123 & 1741 & 4.67\% & 4.50\% \\
3 & 25.0243507 & \multicolumn{1}{l|}{1720} & 22.259355 & 1508 & 11.05\% & 12.33\% \\
4 & 17.1994425 & \multicolumn{1}{l|}{1258} & 16.2014997 & 1217 & 5.80\% & 3.26\% \\
5 & 19.314512 & \multicolumn{1}{l|}{1372} & 18.3067573 & 1245 & 5.22\% & 9.26\%
\end{tabular}
\end{table}

As we can see, all five programs improve their energy efficiency. Our
optimization improves their energy consumption between $4.37\%$ and $11.05\%$, with an average of $6.2\%$.

\section{Threats}
\label{sec:Threats}
The goal of our experiments was to define the energy
consumption profile of JCF implementations and validate such results. 
As in any experiment, there are a few threats to its validity.
We start by presenting the validity threats for the first experiment,
that is, the evaluation of the energy consumption of several Java data structure methods.
We divide these threats in four
categories as defined in~\cite{cook1979quasi}, namely: conclusion validity, 
internal validity, construct validity, and external validity.
\setcounter{paragraph}{0}
\subsection{JCF Implementations Profile}
We start by discussing the threats to validity of the first experiment.

\paragraph{Conclusion Validity}
We used the energy consumption measurements to establish a simplistic order
between the different implementations.
To do so, we have based ourselves on an existing benchmark
(although developed to measure different things).
To perform the actual measurements, we used RAPL which is known
to be a quite reliable tool~\cite{Hahnel2012,DBLP:journals/micro/RotemNAWR12}.
Thus, we believe the finding are quite reliable.

\paragraph{Internal Validity}
The energy consumption measurements we have for the different implementations/methods 
could have been influenced by other factors other than just their source code execution.
To mitigate this issue, for every test
we added a ``warm-up'' run, and we ran every test 10 times, taking 
the average values for these runs so we could minimize particular
states of the machine and other software in it (e.g. operating system daemons). Moreover, we ran
the tests in a Linux server with no other software running
except for the operating system and its services in order to
isolate the energy consumption values for the code we were running as much as possible.

\paragraph{Construct Validity}
We have designed a set of tests to evaluate the energy consumption 
of the methods of the different JCF implementations. As software engineers ourselves, we
have done the best we can and know to make them as real and interesting as
possible. However, these experiments could have been done in many other different ways.
In particular, we have only used strings to perform our evaluation.
We have also fixed the size of the collections in 25K, 250K, and 1M.
Nevertheless, we believe that since all the tests are the same
for all the implementations (of a particular interface), different tests would probably produce the
same relationship between the consumption of the different implementations and their methods.
Still, we make all our material publicly available for better
analysis of our peers.

\paragraph{External Validity}
The experiment we performed can easily be extended to include other
collections. The method can also be easily adapted to other programming languages.
However, until such execution are done, nothing can be said about such results.

\setcounter{paragraph}{0}
\subsection{Validating the Measurements}

Next we present the threats to validity, again divided in four categories,
for the experiment we performed to evaluate the impact of the finding of the
first study when changing the implementations in a complete program.

\paragraph{Conclusion Validity}
Our validation assumed that each method is on the same level of importance or weight, 
and does not distinguish between possible gain of optimizing for one method or another 
(for instance, there might be more gain in optimizing for a commonly used add method over a
retainAll method). Thus, the method of choosing the best alternative implementation would need fine tuning.
Nevertheless, it is consistent that changing an implementation by another
influences the energy consumption of the code in the same line
with the results found for the implementations/methods in the first experiment.

\paragraph{Internal Validity}
The energy consumptions measures we have for the different projects 
(before and after changing the used implementations) could have influence from other factors.
However, the most important thing is the relationship between 
the value before and after changing the implementations.
Nevertheless, we have executed each project 10 times and
calculated the average so particular states of the machine could be
mitigated as much as possible in the final results.

\paragraph{Construct Validity}
We used 5
different (project) implementation of a single problem developed by students in the second semester
of an undergraduation in Computer Science.
This gave us different solutions for the same problem that can be directly
compared as they all passed a set of functional tests defined in the corresponding course.
However, different kinds of projects could have different results.
Nevertheless, there is no basis to suspect that these projects are best or worst
than any other kind. Thus, we expect to continue to see
gains/losses when changing implementations in any other kinds of software projects
according to our findings.

\paragraph{External Validity}
The used source code has no particular characteristics which could
influence our findings. The main characteristic is possibly the fact that
it was developed by novice programmers. Nevertheless, we could see the impact
of changing data structure implementations in both the good and bad (project) implementations.
Thus, we believe that these results can be further generalized for other projects.
Nevertheless, we intend to further study this issue and perform a wider
evaluation.

\section{Related Work}
\label{sec:Related}
Although energy consumption analysis is an area explored for the last two decades, only more recently has it started to focus on software improvement more than hardware improvement. In fact, designing energy-aware programming languages is an active area~\cite{cohen2012energy}, and software developers claim for tools and techniques to support them in the goal of developing energy-aware development~\cite{pinto2014mining}. Even in software testing, researchers want to know how to reduce energy consumption and where do they need to focus to do it~\cite{li2014integrated}.

Studies have shown that there are a lot of software development related factors that can significantly influence the energy consumption of a software system. Different design patterns, using Model-View-Controller, information hiding, implementation of persistence layers, code obfuscations, refactorings, and the usage of different data structures~\cite{brandolese2002impact,li2013calculating,linares2014mining,manotas2014seeds,sahin2012initial,sahin2014code,sahindoes, vetro2013definition} can all influence energy consumption, and all are software related implementation decisions.

Some other research works are even focused on detecting excessive/anomalous energy consumptions in software, not by comparing the overall energy consumption of different implementations of the same software system, but by using tools and techniques specialized in determining the consumption per blocks of code, such as methods~\cite{couto2014detecting}, source code instructions~\cite{li2013calculating} or even bytecode instructions~\cite{hao2012greens}. Those works are based on an energy consumption model: a prediction model which can relate such blocks of code with the amount of energy that they are expected to consume. The concept of energy models have been widely used, particularly in the mobile area~\cite{hao2012greens, hao13icse, nakajima2013model, nakajima2015model,pathak2011, zhang2010}.

In a more focused and concrete way, some research works also analyzed the efficiency of data structures~\cite{manotas2014seeds}. Manotas et al.~\cite{manotas2014seeds} built a framework which was capable of determining the gain or loss, in a global point of view, of switching from one Java collection to another. Nevertheless, we have studied the behavior of a broader set of data structure implementations, divided between the appropriate groups, different population sizes, and a larger number of operations per structure. More recently, a study of the energy profiles of java collection classes has been performed~\cite{icse16}. As this paper is currently not published nor public, unfortunately we cannot properly compare the two works. Nevertheless, the energy quantification is not the only and main contribution in our paper.

\section{Future Work}
\label{sec:FutureWork}
There are several directions for future work. We will continue to evolve our data and perform further tests and analyses. More specifically, we will extend our tests to evaluate other population sizes, more data structures, interface specific methods, and other types of input other than Strings. 

To choose the most efficient alternative implementation, we are defining a new algorithm which uses the number of occurrences of the methods, and different weights for different methods.

We are also planning to extend this work into an automatic analysis and refactoring tool plugin. This tool would would detect if an energy inefficient data structure is being used, suggest an alternative energy efficient data structure, and even automatically refactor the source-code to optimize consumption. The chosen alternative would be based on which methods are being used, either automatically chosen or chosen by 

\section{Conclusion}
\label{sec:Conclusions}

This paper presented a detailed study of the energy consumption of the
\Sets, \Lists and \Maps data structures included in the Java collections
framework. We presented a quantification of the energy spent by each
API method of each of those data structures (\textbf{RQ 1} answer).

Moreover, we introduced a very simple methodology to optimize Java
programs. Based on their JCF data structures and methods, and our
energy quantifications, a transformation to decrease the energy
consumption is suggested. We have presented our first experimental
results that show a decrease of $6.2\%$ in energy consumption
(\textbf{RQ 2} answer).

\bibliographystyle{IEEEtran}
\bibliography{green}



\section*{Supplementary material}

# APPENDIX

Pereira, Couto, Cunha, Fernandes, and Saraiva



# Appendices

## A  Set data for 25k population

| Methods | Concurrent SkipListSet | | HashSet | | Linked HashSet | | TreeSet | |
|---|---|---|---|---|---|---|---|---|
| | J | ms | J | ms | J | ms | J | ms |
| add | 1.6822 | 87 | 1.7749 | 87 | 1.4917 | 75 | 1.4817 | 92 |
| addAll | 1.4549 | 93 | 1.4771 | 89 | 1.9335 | 94 | 1.5101 | 93 |
| clear | 1.4901 | 78 | 1.0586 | 64 | 1.3288 | 60 | 1.8566 | 73 |
| contains | 1.4213 | 88 | 2.0685 | 78 | 1.0401 | 76 | 2.0446 | 79 |
| containsAll | 1.8317 | 96 | 1.4000 | 77 | 2.1748 | 88 | 1.4443 | 89 |
| iterateAll | 1.9225 | 99 | 1.4554 | 92 | 1.2907 | 83 | 1.3844 | 83 |
| iterator | 1.6096 | 83 | 1.7596 | 75 | 0.9613 | 76 | 1.7239 | 76 |
| remove | 1.7877 | 78 | 1.2633 | 75 | 1.2458 | 93 | 1.0700 | 76 |
| removeAll | 1.8072 | 85 | 2.1359 | 77 | 1.9145 | 100 | 1.3920 | 91 |
| retainAll | 3.2607 | 206 | 2.4092 | 200 | 2.2512 | 199 | 3.2222 | 193 |
| toArray | 1.4789 | 86 | 1.3833 | 80 | 1.3776 | 79 | 1.6292 | 80 |

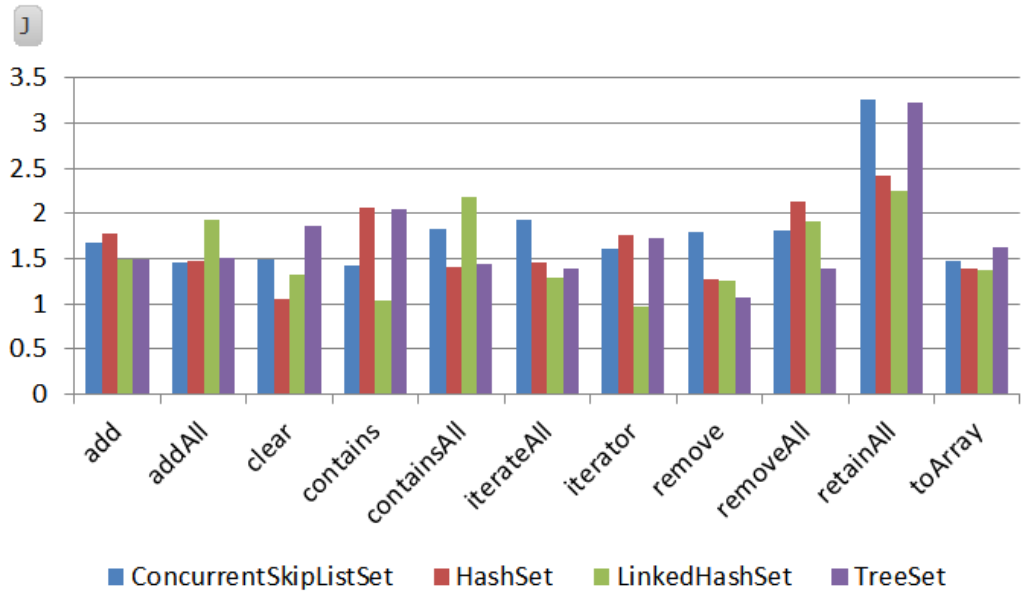

# B  Set data for 250k population

| Methods | Concurrent SkipListSet J | ms | HashSet J | ms | Linked HashSet J | ms | TreeSet J | ms |
|---|---|---|---|---|---|---|---|---|
| add | 5.1254 | 284 | 5.5141 | 305 | 5.2348 | 277 | 4.2959 | 233 |
| addAll | 6.3474 | 391 | 5.5436 | 307 | 5.0125 | 279 | 4.5978 | 288 |
| clear | 4.2285 | 246 | 4.1085 | 227 | 4.6213 | 241 | 4.2033 | 229 |
| contains | 5.0685 | 272 | 5.2959 | 290 | 6.0423 | 251 | 4.5884 | 216 |
| containsAll | 6.5672 | 351 | 5.5917 | 291 | 5.1875 | 259 | 4.4657 | 254 |
| iterateAll | 5.4969 | 266 | 6.2567 | 298 | 5.3597 | 270 | 4.3067 | 247 |
| iterator | 4.9175 | 249 | 5.4548 | 290 | 5.1849 | 265 | 4.1703 | 214 |
| remove | 5.0868 | 260 | 5.3462 | 270 | 4.8186 | 255 | 4.2111 | 225 |
| removeAll | 5.7957 | 364 | 4.9622 | 295 | 5.2345 | 260 | 4.7445 | 278 |
| retainAll | 184.7828 | 17079 | 186.6178 | 17367 | 179.7946 | 16563 | 188.2510 | 17516 |
| toArray | 5.3265 | 270 | 5.7374 | 305 | 4.7959 | 271 | 4.5845 | 245 |

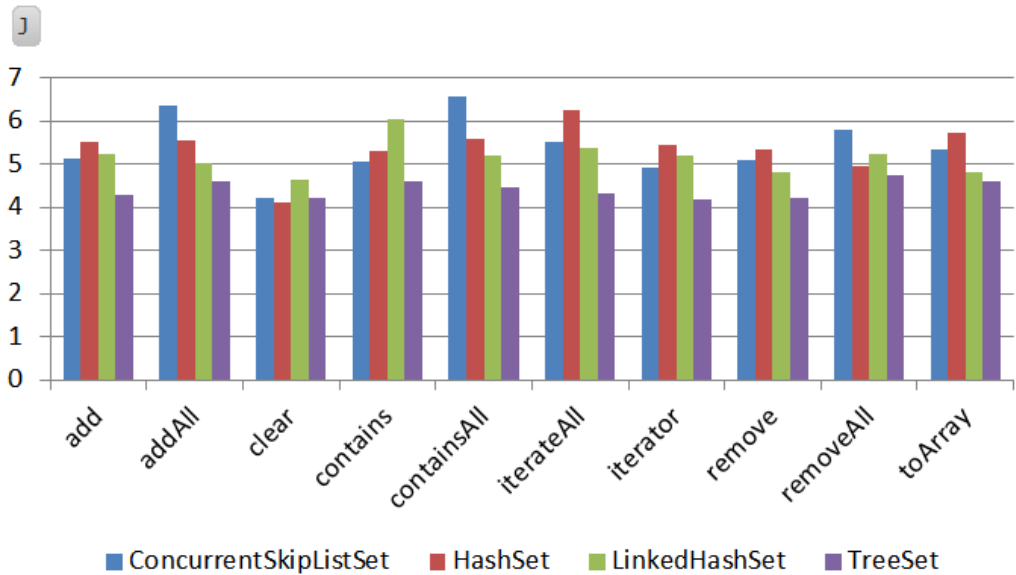

4

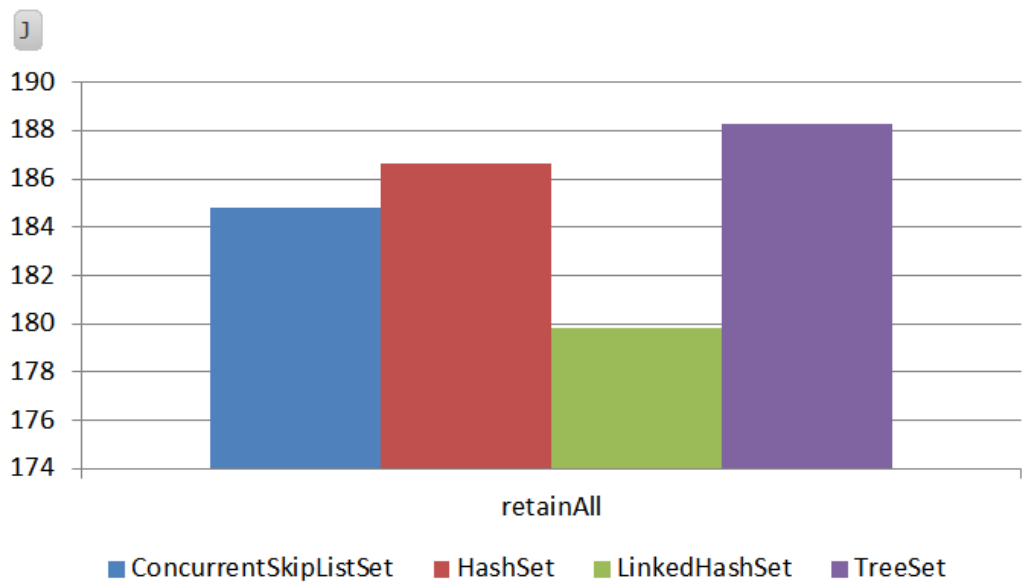

# C  Set data for 1m population

| Methods | Concurrent SkipListSet | | HashSet | | Linked HashSet | | TreeSet | |
|---|---|---|---|---|---|---|---|---|
| | J | ms | J | ms | J | ms | J | ms |
| add | 14.0472 | 824 | 17.5243 | 1072 | 15.0643 | 876 | 14.1021 | 758 |
| addAll | 19.5092 | 1518 | 17.8589 | 1100 | 16.5155 | 983 | 13.5737 | 983 |
| clear | 11.5958 | 747 | 12.0199 | 764 | 12.2874 | 770 | 11.5565 | 758 |
| contains | 13.6576 | 870 | 16.6950 | 1014 | 15.6210 | 880 | 11.2337 | 682 |
| containsAll | 16.9809 | 1212 | 17.2110 | 1038 | 15.8865 | 886 | 12.3979 | 844 |
| iterateAll | 13.0184 | 785 | 18.1706 | 1091 | 15.4155 | 865 | 11.2088 | 684 |
| iterator | 13.2534 | 752 | 16.7433 | 1013 | 15.5284 | 850 | 11.0499 | 641 |
| remove | 12.7444 | 789 | 15.5699 | 949 | 13.6615 | 799 | 11.2653 | 675 |
| removeAll | 17.2849 | 1293 | 17.0514 | 998 | 14.5821 | 841 | 13.2071 | 937 |
| retainAll | 3621.9872 | 346898 | 3912.0129 | 384829 | 3584.3529 | 346337 | 4111.2397 | 408297 |
| toArray | 14.8120 | 875 | 17.8458 | 1070 | 14.3511 | 848 | 13.1271 | 750 |

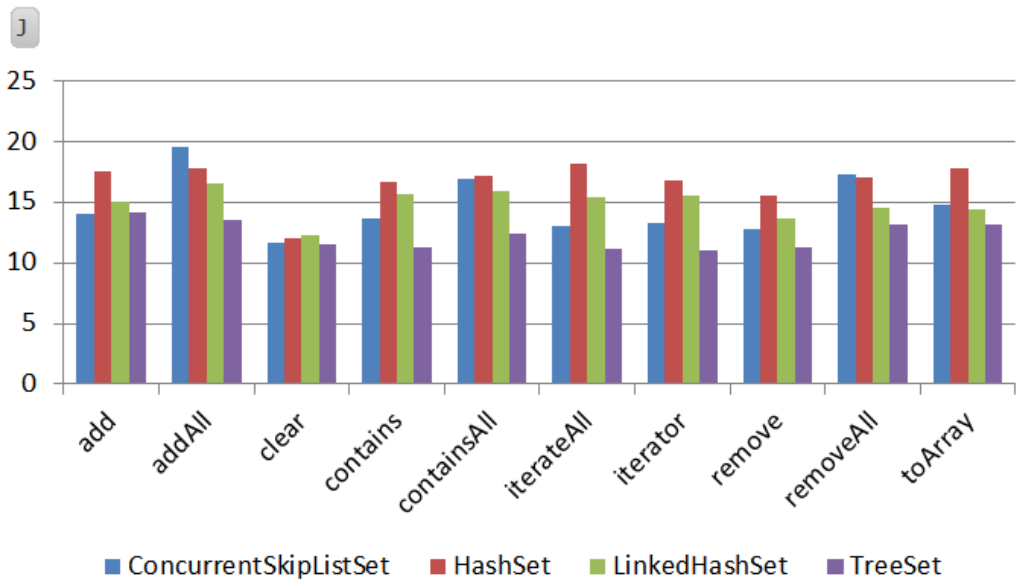

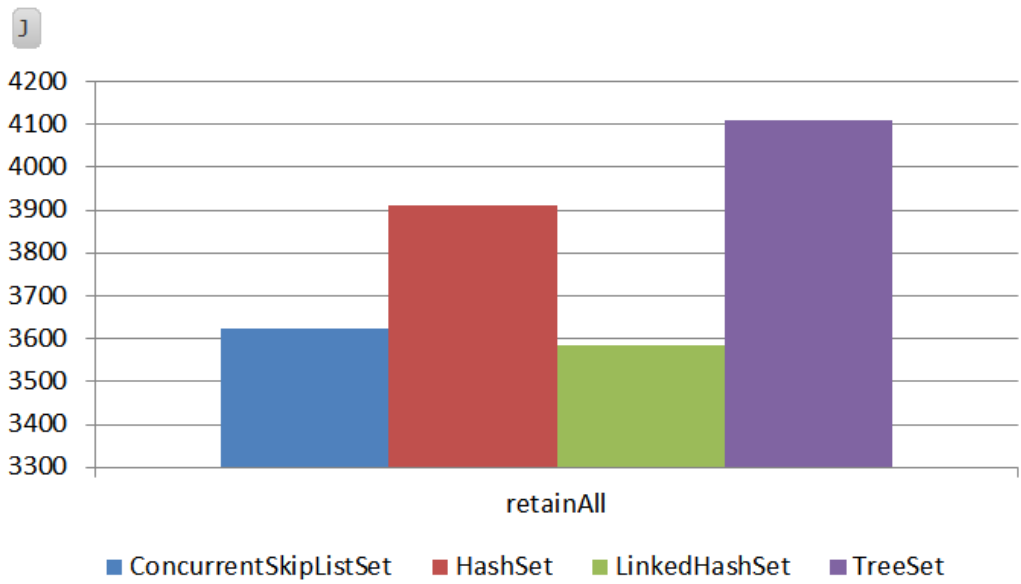

# D List data for 25k population

| Methods | ArrayList J | ms | AttributeList J | ms | CopyOn Write ArrayList J | ms | LinkedList J | ms | RoleList J | ms | Role Unresolved List J | ms | Stack J | ms | Vector J | ms |
|---|---|---|---|---|---|---|---|---|---|---|---|---|---|---|---|---|
| add | 0.9773 | 71 | 1.1510 | 67 | 1.7839 | 117 | 1.8016 | 86 | 1.4801 | 76 | 1.1865 | 74 | 1.5659 | 76 | 1.5177 | 69 |
| addAll | 1.3353 | 76 | 1.0492 | 88 | 1.3586 | 82 | 1.1043 | 88 | 1.6661 | 76 | 1.8672 | 88 | 1.1015 | 88 | 1.7903 | 73 |
| addAlli | 1.7855 | 86 | 1.6035 | 68 | 1.1789 | 86 | 1.7272 | 99 | 1.5980 | 81 | 1.2497 | 85 | 1.2962 | 72 | 1.6268 | 90 |
| addI | 1.7125 | 93 | 1.3849 | 87 | 1.6558 | 119 | 1.6404 | 96 | 1.2718 | 85 | 1.3124 | 86 | 1.5287 | 83 | 1.4554 | 86 |
| clear | 1.1284 | 76 | 1.2409 | 75 | 1.7155 | 68 | 1.6497 | 74 | 1.6705 | 76 | 1.4304 | 80 | 1.6199 | 73 | 1.0574 | 71 |
| contains | 2.7568 | 166 | 2.4228 | 165 | 3.1768 | 167 | 3.1552 | 193 | 2.1751 | 162 | 2.4688 | 164 | 2.0128 | 166 | 2.1558 | 168 |
| containsAll | 1.5993 | 87 | 1.8053 | 92 | 2.1889 | 92 | 2.2887 | 118 | 1.3244 | 100 | 1.3930 | 96 | 1.2054 | 89 | 1.5091 | 87 |
| get | 2.0029 | 83 | 1.1171 | 78 | 1.4918 | 77 | 2.0168 | 109 | 2.2110 | 81 | 1.6613 | 71 | 1.8956 | 86 | 1.4978 | 73 |
| indexOf | 1.4447 | 76 | 2.0325 | 84 | 1.5682 | 70 | 2.6289 | 101 | 1.5674 | 79 | 1.1944 | 81 | 1.8090 | 81 | 2.0788 | 75 |
| iterateAll | 2.0701 | 79 | 1.0473 | 77 | 1.0103 | 73 | 2.6401 | 107 | 1.3605 | 85 | 1.7822 | 71 | 1.6036 | 81 | 1.1336 | 87 |
| iterator | 1.4893 | 84 | 1.1589 | 84 | 1.3922 | 72 | 1.7666 | 108 | 1.9760 | 73 | 1.3300 | 79 | 2.1895 | 84 | 1.6505 | 83 |
| lastIndexOf | 1.7750 | 99 | 1.7666 | 98 | 2.0383 | 94 | 2.5019 | 127 | 1.8914 | 92 | 1.4211 | 95 | 1.2260 | 84 | 1.2296 | 96 |
| listIterator | 1.4457 | 76 | 1.6190 | 84 | 1.3737 | 71 | 2.5003 | 106 | 1.3380 | 80 | 1.5176 | 85 | 1.6354 | 69 | 1.2746 | 81 |
| listIteratori | 1.7356 | 78 | 1.1552 | 81 | 1.5160 | 77 | 2.1996 | 105 | 1.7588 | 79 | 1.0334 | 80 | 1.8799 | 85 | 1.7545 | 78 |
| remove | 1.1308 | 96 | 1.4480 | 85 | 2.1946 | 162 | 1.6924 | 98 | 1.4560 | 84 | 1.1368 | 85 | 1.2663 | 96 | 1.4973 | 82 |
| removeAll | 8.0905 | 671 | 7.8108 | 697 | 7.3237 | 666 | 8.3150 | 752 | 7.6148 | 692 | 7.9911 | 664 | 7.3824 | 654 | 7.1281 | 665 |
| removei | 1.9135 | 85 | 1.3534 | 92 | 2.2858 | 118 | 1.7174 | 100 | 1.6308 | 85 | 1.6369 | 89 | 1.5850 | 81 | 1.5486 | 90 |
| retainAll | 2.7037 | 193 | 2.7845 | 200 | 2.6052 | 198 | 2.5982 | 205 | 3.0973 | 197 | 2.4172 | 200 | 2.7635 | 242 | 3.4019 | 245 |
| set | 0.9476 | 64 | 1.5943 | 70 | 1.9669 | 110 | 2.0474 | 112 | 1.5249 | 76 | 1.2312 | 73 | 1.4938 | 75 | 1.4957 | 72 |
| sublist | 1.3108 | 76 | 1.6021 | 80 | 1.4792 | 80 | 1.8457 | 98 | 1.4910 | 85 | 1.5117 | 71 | 1.7082 | 75 | 0.9414 | 75 |
| toArray | 1.6418 | 84 | 1.5024 | 84 | 2.0934 | 73 | 1.6739 | 106 | 1.5418 | 79 | 1.7455 | 83 | 1.5694 | 69 | 2.0213 | 80 |

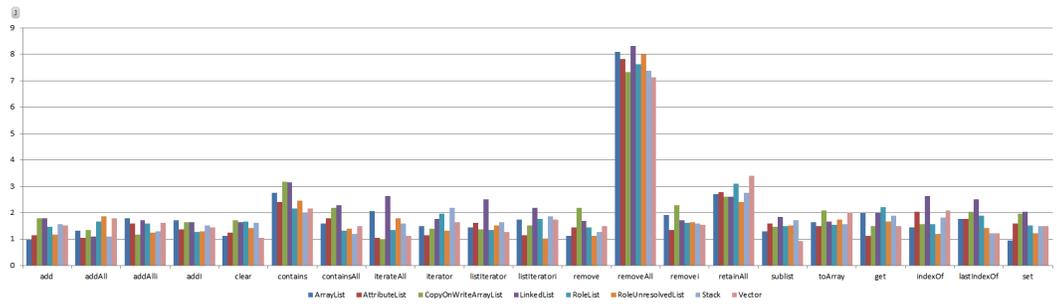

8

# E List data for 250k population

| Methods | ArrayList | | AttributeList | | CopyOnWrite ArrayList | | LinkedList | | RoleList | | Role Unresolved List | | Stack | | Vector | |
|---|---|---|---|---|---|---|---|---|---|---|---|---|---|---|---|---|
| | J | ms | J | ms | J | ms | J | ms | J | ms | J | ms | J | ms | J | ms |
| add | 3.8352 | 225 | 4.6749 | 228 | 59.9706 | 5552 | 6.2024 | 294 | 4.0685 | 223 | 4.3435 | 222 | 4.3659 | 225 | 4.0186 | 221 |
| addAll | 4.0158 | 213 | 4.1970 | 213 | 3.7459 | 212 | 6.3524 | 306 | 4.0595 | 236 | 4.1118 | 233 | 4.0586 | 211 | 3.8572 | 217 |
| addAlli | 3.1822 | 200 | 3.6327 | 196 | 3.6868 | 195 | 6.5870 | 318 | 3.8740 | 221 | 4.3522 | 198 | 4.3606 | 222 | 3.9313 | 229 |
| addI | 17.9630 | 1366 | 17.5091 | 1372 | 64.4550 | 6091 | 13.8804 | 1037 | 17.3695 | 1371 | 17.1374 | 1366 | 17.4695 | 1365 | 17.4099 | 1368 |
| clear | 3.6320 | 218 | 4.2446 | 238 | 4.1154 | 219 | 4.2423 | 225 | 4.3240 | 221 | 4.0848 | 237 | 4.1995 | 220 | 4.4717 | 238 |
| contains | 149.2101 | 14930 | 143.8105 | 14297 | 138.7096 | 13608 | 177.4864 | 17145 | 143.5583 | 14361 | 144.8360 | 14440 | 148.3477 | 14941 | 147.6217 | 14919 |
| containsAll | 14.2118 | 1210 | 14.1992 | 1161 | 15.4383 | 1282 | 16.9662 | 1399 | 13.9955 | 1178 | 14.3545 | 1184 | 14.0874 | 1155 | 14.2835 | 1160 |
| get | 3.7491 | 225 | 3.8010 | 196 | 3.3272 | 196 | 9.1736 | 738 | 4.1234 | 194 | 3.3873 | 203 | 4.1009 | 200 | 3.4824 | 195 |
| indexOf | 4.2838 | 207 | 4.2417 | 208 | 4.0170 | 213 | 5.5329 | 275 | 4.1047 | 230 | 3.6321 | 201 | 3.6502 | 202 | 4.0763 | 204 |
| iterateAll | 3.9417 | 205 | 3.8646 | 206 | 4.1047 | 200 | 6.0103 | 303 | 3.9434 | 229 | 4.3596 | 209 | 3.9510 | 224 | 4.0285 | 232 |
| iterator | 4.2709 | 196 | 4.2472 | 194 | 3.7052 | 200 | 6.2585 | 266 | 3.7146 | 201 | 3.7738 | 204 | 3.4071 | 204 | 4.2383 | 204 |
| lastIndexOf | 27.2224 | 2438 | 26.7242 | 2408 | 26.5297 | 2448 | 36.6855 | 3260 | 26.6724 | 2413 | 26.8274 | 2388 | 26.2273 | 2409 | 26.5668 | 2413 |
| listIterator | 4.0923 | 199 | 3.8791 | 197 | 3.4499 | 201 | 4.7948 | 271 | 3.5211 | 198 | 3.5665 | 197 | 3.5485 | 195 | 3.8970 | 200 |
| listIteratori | 3.5178 | 197 | 4.8897 | 195 | 4.1712 | 197 | 5.6153 | 264 | 4.1498 | 195 | 3.7611 | 196 | 3.4433 | 200 | 3.8707 | 195 |
| remove | 16.7118 | 1293 | 16.0287 | 1296 | 85.5149 | 8048 | 5.2105 | 243 | 16.3957 | 1305 | 16.9656 | 1294 | 15.8466 | 1297 | 16.4117 | 1290 |
| removeAll | 800.6362 | 74570 | 815.9125 | 75055 | 848.7941 | 77949 | 829.8120 | 76059 | 812.6019 | 75188 | 803.4346 | 74341 | 811.8826 | 75083 | 816.0893 | 75628 |
| removei | 15.7604 | 1238 | 15.5615 | 1238 | 57.5978 | 5359 | 9.4499 | 731 | 15.3716 | 1232 | 15.9824 | 1235 | 15.9731 | 1225 | 15.4837 | 1243 |
| retainAll | 191.2629 | 17515 | 189.2622 | 17436 | 187.5156 | 17258 | 189.0232 | 17346 | 185.6256 | 17141 | 192.1342 | 17445 | 246.8980 | 22740 | 256.3624 | 23843 |
| set | 4.0179 | 204 | 3.8355 | 207 | 55.3024 | 5215 | 9.4216 | 757 | 3.9253 | 211 | 3.8588 | 210 | 4.1395 | 208 | 4.2365 | 236 |
| sublist | 4.1009 | 203 | 3.4186 | 202 | 3.5210 | 194 | 5.3223 | 272 | 3.6955 | 197 | 3.9814 | 196 | 3.7111 | 194 | 3.9882 | 198 |
| toArray | 3.6473 | 200 | 3.8151 | 202 | 3.5124 | 196 | 6.5844 | 273 | 3.8092 | 195 | 3.7162 | 196 | 3.7170 | 195 | 3.1472 | 195 |

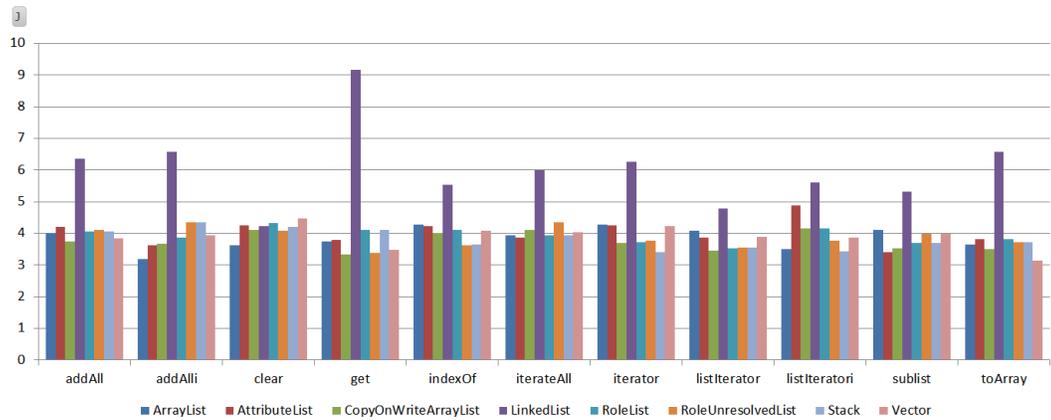

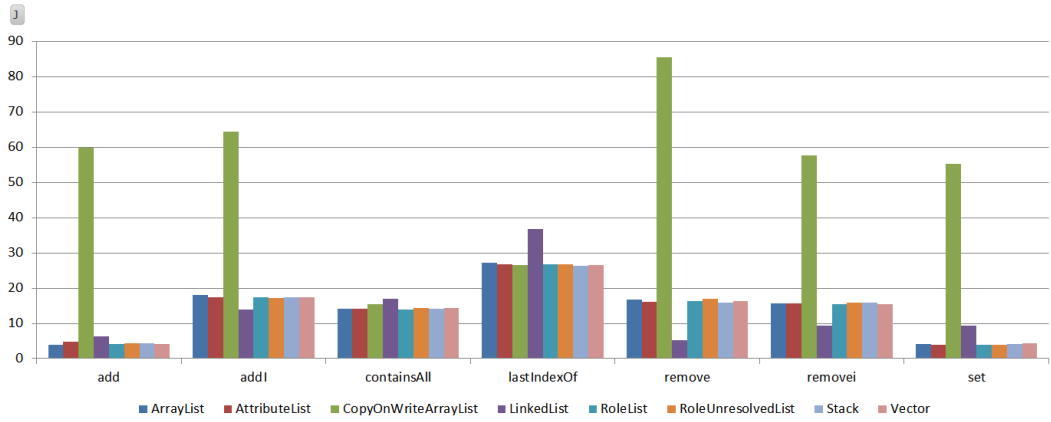

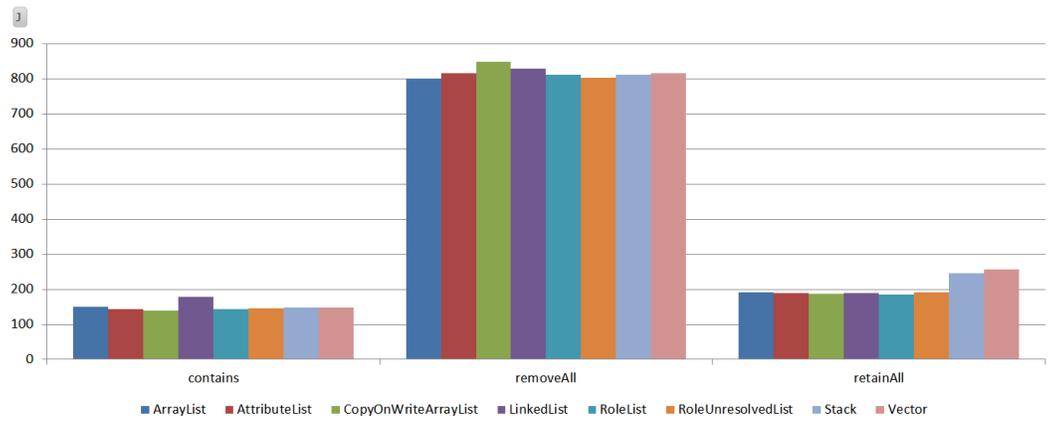

# F   List data for 1m population

| | ArrayList | | AttributeList | | CopyOn Write | | LinkedList | | RoleList | | Role Unresolved | | Stack | | Vector | |
|---|---|---|---|---|---|---|---|---|---|---|---|---|---|---|---|---|
| Methods | J | ms | J | ms | J | ms | J | ms | J | ms | J | ms | J | ms | J | ms |
| add | 12.0464 | 701 | 12.7195 | 707 | 1192.1624 | 116076 | 13.1935 | 728 | 12.4607 | 709 | 12.3262 | 706 | 11.7879 | 722 | 11.3576 | 696 |
| addAll | 10.9094 | 648 | 11.3724 | 641 | 11.9216 | 660 | 14.2490 | 817 | 11.3734 | 637 | 11.4211 | 639 | 10.4004 | 642 | 10.7678 | 649 |
| addAlli | 10.1839 | 615 | 10.5913 | 622 | 10.5091 | 634 | 14.0827 | 834 | 10.4132 | 613 | 10.0168 | 611 | 10.0941 | 615 | 10.5229 | 617 |
| addI | 308.6728 | 28059 | 308.4402 | 28070 | 1347.1859 | 128466 | 190.3535 | 18902 | 312.4176 | 28542 | 313.0971 | 28577 | 312.1597 | 28570 | 312.8684 | 28688 |
| clear | 11.6159 | 740 | 11.7247 | 730 | 12.0538 | 732 | 12.0858 | 742 | 11.6814 | 738 | 11.7761 | 736 | 12.2983 | 736 | 11.7146 | 736 |
| contains | 1970.9064 | 195324 | 1952.7448 | 191984 | 1979.2615 | 195337 | 2466.3537 | 239826 | 1950.6017 | 191656 | 1959.7723 | 193343 | 1957.1560 | 192235 | 1968.4488 | 193401 |
| containsAll | 185.4605 | 17164 | 185.8306 | 17224 | 199.4027 | 18557 | 270.4051 | 24800 | 185.9840 | 17178 | 186.6167 | 17244 | 187.1497 | 17239 | 186.1992 | 17221 |
| get | 10.9069 | 616 | 11.4765 | 625 | 10.7958 | 629 | 81.1299 | 8346 | 11.3632 | 643 | 11.0016 | 626 | 10.9420 | 621 | 10.7961 | 616 |
| indexOf | 12.3355 | 752 | 12.5442 | 752 | 12.4852 | 751 | 13.0304 | 824 | 12.5427 | 758 | 11.8597 | 750 | 12.6218 | 771 | 12.1569 | 764 |
| iterateAll | 11.4806 | 659 | 11.3767 | 646 | 10.8094 | 646 | 12.0847 | 692 | 10.6485 | 640 | 11.0305 | 644 | 11.9679 | 741 | 11.3126 | 737 |
| iterator | 11.0537 | 635 | 10.9354 | 629 | 10.6816 | 624 | 11.9186 | 694 | 10.5633 | 633 | 10.0707 | 622 | 10.9189 | 653 | 11.3553 | 643 |
| lastIndexOf | 387.2730 | 36766 | 388.6052 | 37032 | 387.0817 | 37231 | 489.1250 | 46240 | 393.5686 | 37582 | 390.2053 | 37104 | 386.6704 | 37474 | 386.9537 | 37475 |
| listIterator | 10.6885 | 611 | 11.4602 | 631 | 11.4474 | 657 | 12.0396 | 669 | 11.3297 | 629 | 11.2530 | 630 | 11.3042 | 623 | 10.8957 | 615 |
| listIteratori | 11.5547 | 640 | 11.8432 | 618 | 10.8368 | 629 | 12.1197 | 741 | 11.1662 | 622 | 11.3287 | 643 | 10.8977 | 629 | 10.9616 | 611 |
| remove | 286.8151 | 25810 | 289.7602 | 26253 | 1382.3727 | 130720 | 11.2295 | 639 | 290.6235 | 26322 | 288.9534 | 26146 | 288.1555 | 25858 | 290.4005 | 26330 |
| removeAll | 6240.3017 | 616088 | 6254.3095 | 619221 | 7633.4344 | 756859 | 6031.4099 | 600757 | 6240.6818 | 616568 | 6242.6897 | 616035 | 7481.8537 | 741031 | 7491.7295 | 740725 |
| removei | 264.0750 | 23596 | 264.7749 | 23789 | 1193.7827 | 113704 | 80.0698 | 8312 | 267.6469 | 24018 | 263.0561 | 23579 | 266.6137 | 23979 | 265.2319 | 23828 |
| retainAll | 4165.0922 | 415147 | 4108.4192 | 409992 | 4285.7077 | 426197 | 4046.2904 | 402778 | 4125.8655 | 411737 | 4143.7322 | 413506 | 6116.5527 | 605233 | 6083.4340 | 606689 |
| set | 11.5278 | 653 | 10.7694 | 647 | 1107.5213 | 107726 | 81.0687 | 8381 | 11.5072 | 655 | 11.8232 | 642 | 11.4580 | 673 | 11.1637 | 656 |
| sublist | 10.9601 | 611 | 10.7711 | 636 | 11.1858 | 622 | 11.7203 | 676 | 10.9904 | 638 | 10.8467 | 634 | 11.3011 | 636 | 10.6396 | 627 |
| toArray | 10.7242 | 612 | 11.5362 | 642 | 10.4389 | 628 | 14.5166 | 796 | 10.6314 | 623 | 10.6248 | 619 | 10.6683 | 614 | 11.0372 | 637 |

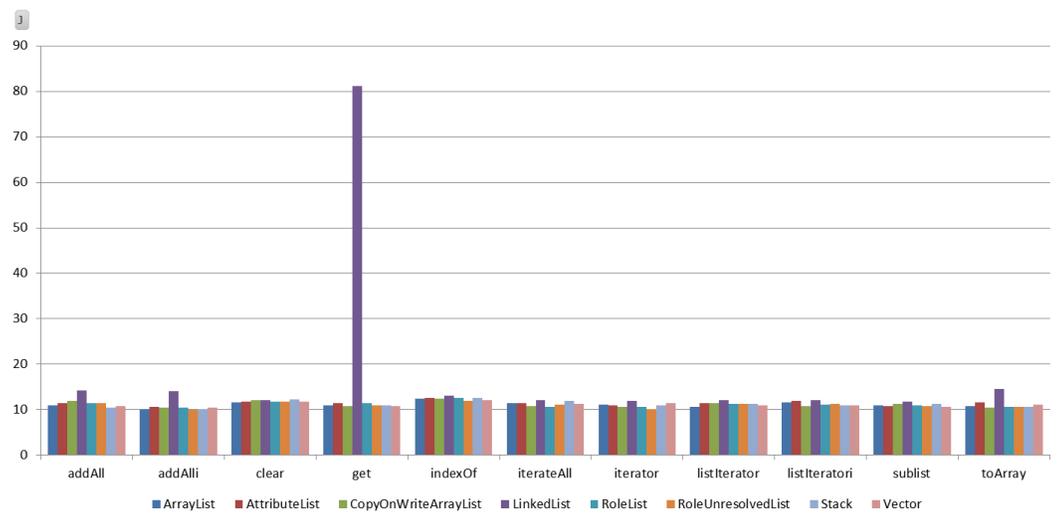

11

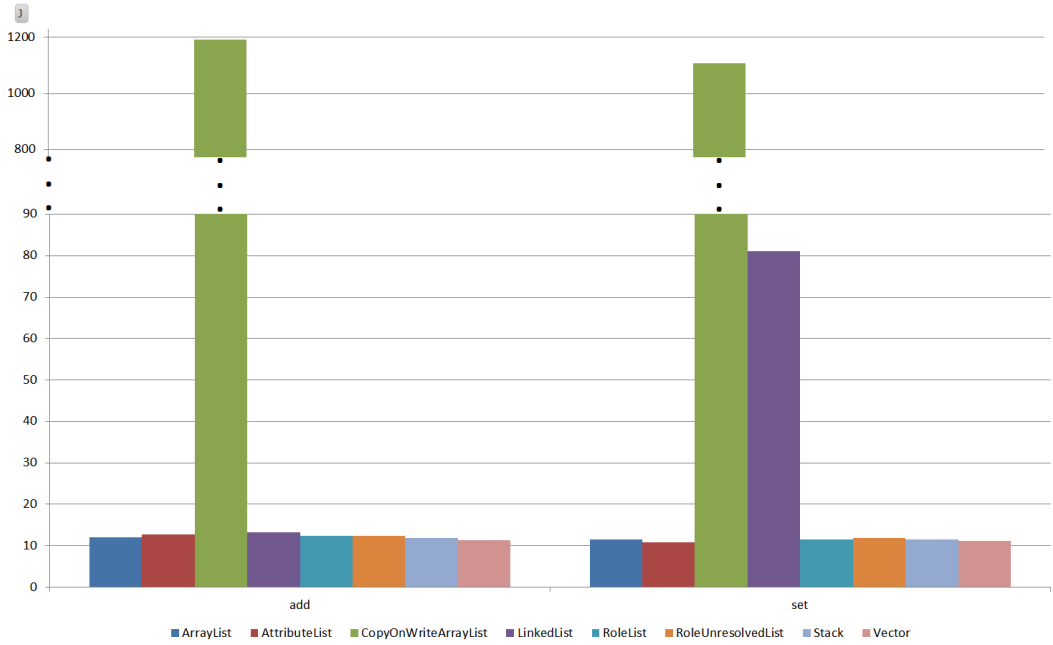

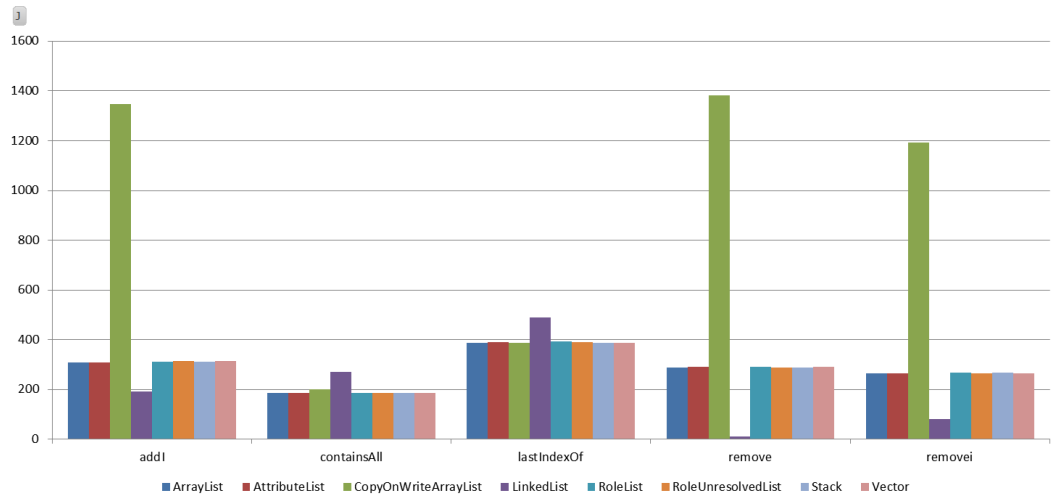

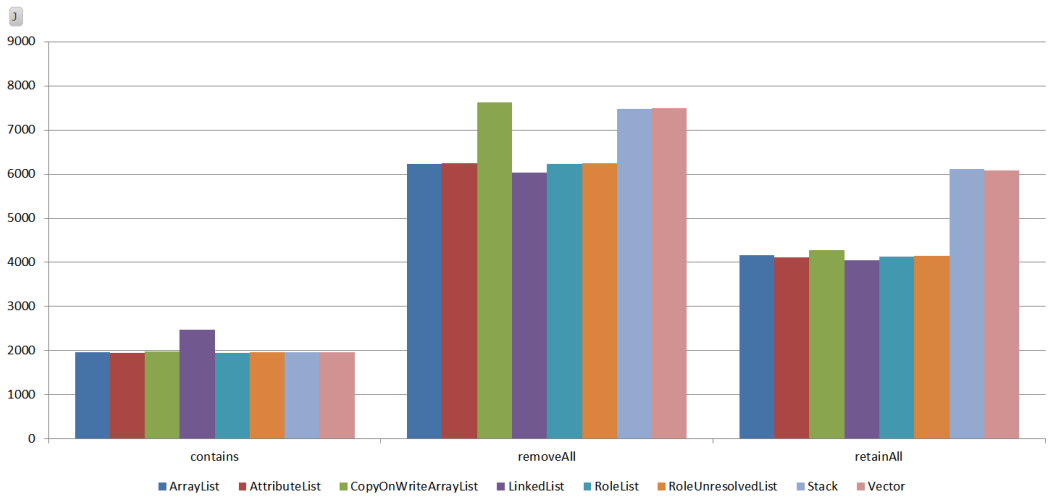

# G Map data for 25k population

| Methods | ConcurrentHashMap J | ms | ConcurrentSkipListMap J | ms | HashMap J | ms | Hashtable J | ms | LinkedHashMap J | ms | Properties J | ms | SimpleBindings J | ms | TreeMap J | ms | UIDefaults J | ms | WeakHashMap J | ms |
|---|---|---|---|---|---|---|---|---|---|---|---|---|---|---|---|---|---|---|---|---|
| clear | 2.0276 | 94 | 2.2961 | 88 | 1.8395 | 104 | 1.5761 | 94 | 1.5025 | 97 | 2.0777 | 98 | 2.1401 | 106 | 1.6706 | 98 | 1.8143 | 105 | 1.9941 | 95 |
| containsKey | 2.3132 | 105 | 2.1693 | 123 | 2.1343 | 103 | 1.8582 | 94 | 1.8726 | 103 | 1.6018 | 107 | 1.8055 | 99 | 1.9452 | 100 | 2.3366 | 89 | 1.9675 | 108 |
| containsValue | 21.5611 | 2305 | 7.8032 | 643 | 8.3615 | 683 | 8.4957 | 765 | 6.1326 | 462 | 7.3755 | 692 | 7.9912 | 678 | 9.1771 | 847 | 7.9341 | 714 | 6.7072 | 562 |
| entrySet | 2.2878 | 93 | 2.2363 | 116 | 1.8531 | 108 | 2.1332 | 107 | 1.8362 | 113 | 1.7800 | 97 | 2.1557 | 102 | 2.1617 | 115 | 1.7087 | 105 | 1.4666 | 102 |
| get | 2.3106 | 103 | 1.9972 | 119 | 1.8120 | 102 | 1.4071 | 100 | 1.8252 | 116 | 1.7851 | 97 | 1.5359 | 100 | 2.2331 | 115 | 1.5252 | 89 | 1.7185 | 103 |
| iterateAll | 2.1041 | 96 | 1.8353 | 115 | 2.6673 | 100 | 1.5343 | 91 | 1.6462 | 111 | 1.6362 | 100 | 2.0472 | 116 | 1.9122 | 111 | 1.6574 | 95 | 1.7139 | 106 |
| keySet | 1.7287 | 95 | 2.4889 | 124 | 1.6813 | 114 | 2.2226 | 99 | 1.8328 | 103 | 1.4866 | 92 | 2.0630 | 106 | 2.1680 | 110 | 1.5547 | 99 | 1.8749 | 105 |
| put | 1.8591 | 104 | 2.2888 | 102 | 2.4628 | 92 | 1.3123 | 96 | 2.0338 | 108 | 1.7038 | 107 | 2.1646 | 102 | 1.4355 | 91 | 2.1204 | 93 | 2.5784 | 105 |
| putAll | 1.4147 | 95 | 2.2852 | 122 | 1.7564 | 100 | 1.5949 | 105 | 1.8608 | 113 | 1.3097 | 95 | 2.1461 | 112 | 1.8914 | 116 | 2.3094 | 87 | 2.0750 | 108 |
| remove | 1.8574 | 92 | 2.2131 | 105 | 1.9256 | 109 | 1.6067 | 97 | 2.2300 | 106 | 1.9660 | 98 | 2.2178 | 106 | 1.8133 | 101 | 1.6888 | 92 | 2.4103 | 103 |
| values | 1.8279 | 85 | 2.4690 | 116 | 2.5755 | 109 | 2.2266 | 94 | 2.0009 | 107 | 1.9120 | 111 | 2.0692 | 108 | 1.4467 | 105 | 1.6533 | 100 | 2.4628 | 111 |

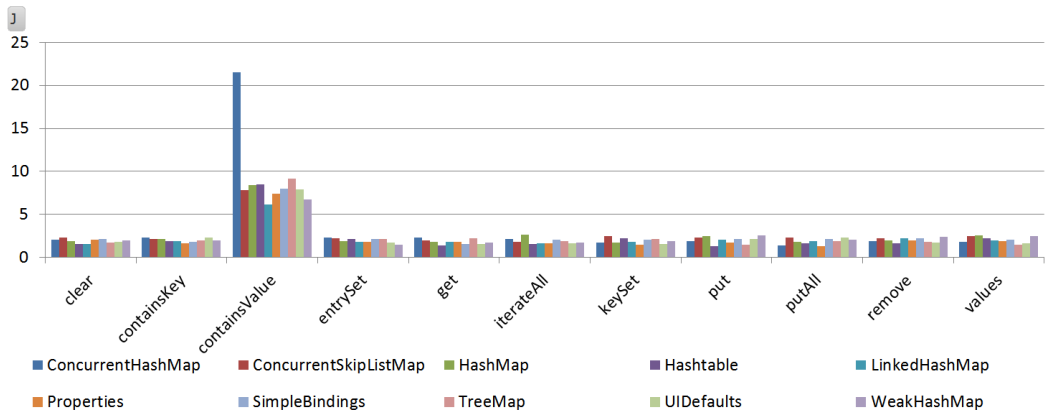

# H Map data for 250k population

| Methods | Concurrent HashMap J | ms | Concurrent SkipList Map J | ms | HashMap J | ms | Hashtable J | ms | Linked HashMap J | ms | Properties J | ms | Simple Bindings J | ms | TreeMap J | ms | UIDefaults J | ms | Weak HashMap J | ms |
|---|---|---|---|---|---|---|---|---|---|---|---|---|---|---|---|---|---|---|---|---|
| clear | 8.0929 | 532 | 8.6506 | 548 | 8.2128 | 526 | 8.4732 | 522 | 7.9924 | 518 | 8.0495 | 529 | 8.4215 | 516 | 8.5468 | 543 | 8.4711 | 527 | 8.1116 | 522 |
| containsKey | 7.8434 | 426 | 9.1539 | 473 | 9.1804 | 550 | 7.8745 | 451 | 7.8521 | 429 | 8.0387 | 451 | 9.1259 | 536 | 7.1873 | 415 | 7.9168 | 452 | 10.8584 | 610 |
| containsValue | 3911.4124 | 461730 | 1334.8256 | 154701 | 730.9166 | 75444 | 1128.1868 | 127397 | 545.2480 | 55788 | 1166.8370 | 131071 | 745.1696 | 76840 | 1606.2652 | 183898 | 1204.5573 | 136323 | 912.0360 | 91357 |
| entrySet | 8.0325 | 434 | 8.0541 | 466 | 9.0436 | 547 | 7.8135 | 438 | 7.8607 | 441 | 7.7498 | 442 | 8.6253 | 540 | 7.3000 | 399 | 8.7348 | 440 | 10.3130 | 621 |
| get | 7.4986 | 434 | 8.1773 | 465 | 9.4016 | 546 | 8.2263 | 449 | 8.0936 | 419 | 8.1691 | 453 | 9.3113 | 542 | 7.2236 | 415 | 8.8114 | 442 | 10.4985 | 621 |
| iterateAll | 7.5645 | 487 | 8.3979 | 494 | 9.2449 | 570 | 8.9272 | 477 | 7.8789 | 451 | 8.5419 | 477 | 9.7741 | 562 | 7.3954 | 452 | 8.5096 | 475 | 10.4295 | 630 |
| keySet | 7.3338 | 436 | 8.2838 | 456 | 9.4461 | 570 | 8.0567 | 448 | 7.5861 | 433 | 8.0351 | 452 | 9.3002 | 533 | 7.5795 | 415 | 7.8576 | 455 | 9.9974 | 614 |
| put | 8.0530 | 458 | 8.6610 | 467 | 9.5894 | 576 | 8.4029 | 452 | 8.3766 | 432 | 8.3302 | 448 | 9.6672 | 557 | 8.2926 | 421 | 7.9345 | 450 | 11.2110 | 633 |
| putAll | 7.9165 | 460 | 9.6607 | 566 | 9.6678 | 562 | 7.4861 | 443 | 7.6870 | 421 | 7.6266 | 441 | 9.3364 | 550 | 7.5239 | 470 | 7.9134 | 451 | 9.7825 | 626 |
| remove | 8.8059 | 455 | 8.0675 | 474 | 9.1076 | 545 | 7.9131 | 437 | 8.4532 | 422 | 8.1008 | 443 | 8.3784 | 531 | 8.0151 | 437 | 7.7442 | 447 | 8.4500 | 533 |
| values | 7.3602 | 440 | 8.7056 | 463 | 9.4056 | 571 | 7.9879 | 450 | 7.8246 | 438 | 8.4208 | 441 | 8.9347 | 548 | 7.3330 | 415 | 8.4531 | 451 | 10.5991 | 615 |

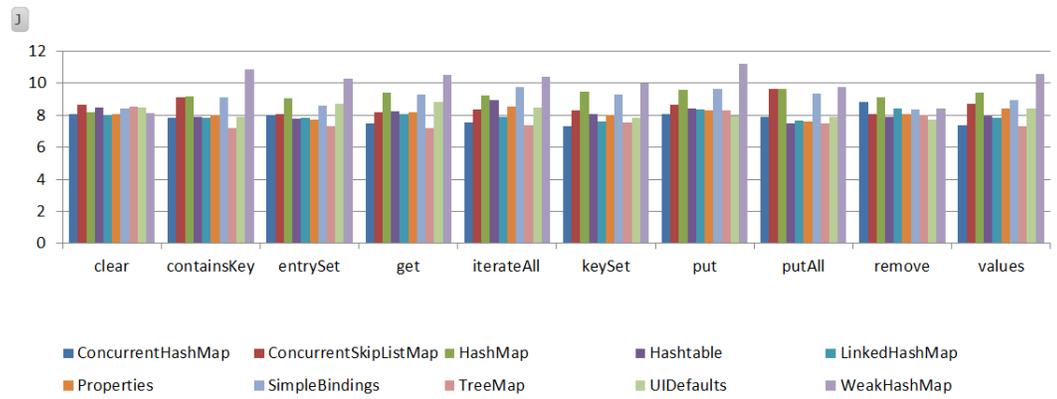

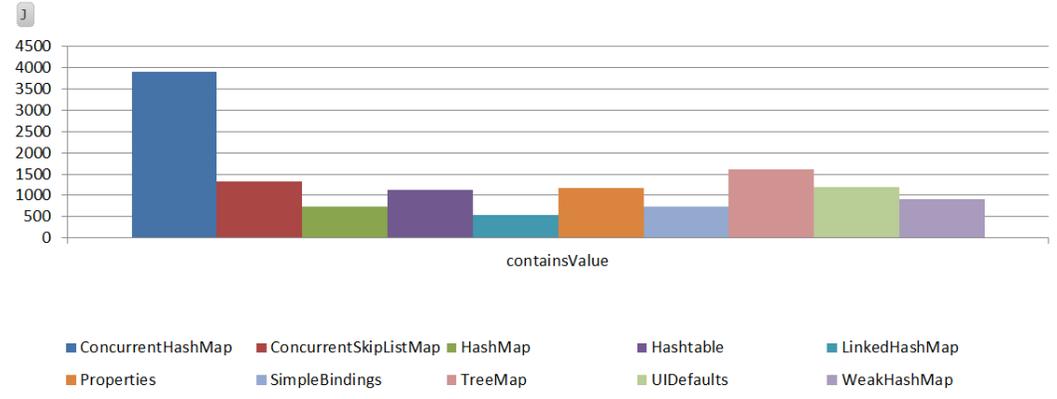

15

# I Map data for 1m population

| Methods | ConcurrentHashMap | | ConcurrentSkipListMap | | HashMap | | Hashtable | | LinkedHashMap | | Properties | | SimpleBindings | | TreeMap | | UIDefaults | | WeakHashMap | |
|---|---|---|---|---|---|---|---|---|---|---|---|---|---|---|---|---|---|---|---|---|
| | J | ms | J | ms | J | ms | J | ms | J | ms | J | ms | J | ms | J | ms | J | ms | J | ms |
| clear | 32.1468 | 2086 | 31.6883 | 2020 | 31.6585 | 1997 | 31.4659 | 2009 | 31.7548 | 2002 | 31.6198 | 2018 | 31.3525 | 1978 | 32.2024 | 2015 | 31.6685 | 2018 | 31.7730 | 1992 |
| containsKey | 25.3230 | 1645 | 27.0269 | 1707 | 33.4752 | 2056 | 27.0329 | 1615 | 26.1077 | 1585 | 26.8887 | 1634 | 32.9557 | 2026 | 24.2985 | 1539 | 26.4794 | 1615 | 38.1398 | 2362 |
| containsValue | 5110.1689 | 602187 | 5223.6619 | 602035 | 5809.9971 | 602007 | 5377.3805 | 602040 | 5719.4071 | 602019 | 5408.3341 | 602026 | 5781.2350 | 602023 | 5291.3279 | 602031 | 5454.6906 | 602028 | 5931.2509 | 602055 |
| entrySet | 26.2053 | 1631 | 26.3433 | 1630 | 34.2838 | 2108 | 27.4266 | 1637 | 27.3013 | 1625 | 27.1060 | 1621 | 32.9119 | 2043 | 23.8757 | 1478 | 26.7149 | 1624 | 38.6346 | 2417 |
| get | 25.4475 | 1658 | 26.8225 | 1668 | 33.0373 | 2060 | 26.7896 | 1603 | 27.0924 | 1605 | 28.3008 | 1683 | 33.4063 | 2061 | 25.0432 | 1545 | 26.9511 | 1601 | 38.2429 | 2357 |
| iterateAll | 29.0126 | 1931 | 26.4318 | 1701 | 34.3681 | 2158 | 27.3898 | 1701 | 27.3926 | 1655 | 27.9918 | 1717 | 33.7128 | 2100 | 24.8704 | 1627 | 28.1027 | 1702 | 34.2753 | 2250 |
| keySet | 26.4697 | 1654 | 26.1301 | 1613 | 33.2621 | 2041 | 27.0295 | 1634 | 27.3374 | 1638 | 26.3382 | 1621 | 33.7798 | 2055 | 24.1255 | 1502 | 27.0168 | 1618 | 38.2719 | 2376 |
| put | 27.0194 | 1759 | 28.0145 | 1745 | 34.9660 | 2168 | 28.3543 | 1712 | 26.1753 | 1552 | 28.0086 | 1731 | 34.3807 | 2141 | 25.3354 | 1587 | 28.5027 | 1712 | 36.0024 | 2211 |
| putAll | 26.5031 | 1764 | 29.3885 | 2020 | 35.0481 | 2174 | 28.0358 | 1704 | 26.2283 | 1580 | 27.5977 | 1682 | 34.2556 | 2121 | 26.6613 | 1725 | 27.4983 | 1673 | 38.2742 | 2380 |
| remove | 25.5665 | 1667 | 26.6545 | 1694 | 32.3202 | 1973 | 26.4978 | 1591 | 25.9951 | 1581 | 26.2373 | 1598 | 31.6204 | 1937 | 24.4253 | 1530 | 27.1669 | 1612 | 35.4727 | 2158 |
| values | 26.3867 | 1663 | 26.6376 | 1632 | 32.8656 | 2033 | 26.9196 | 1630 | 28.2103 | 1628 | 26.8990 | 1621 | 33.3910 | 2052 | 24.2967 | 1477 | 26.7752 | 1639 | 38.9120 | 2413 |

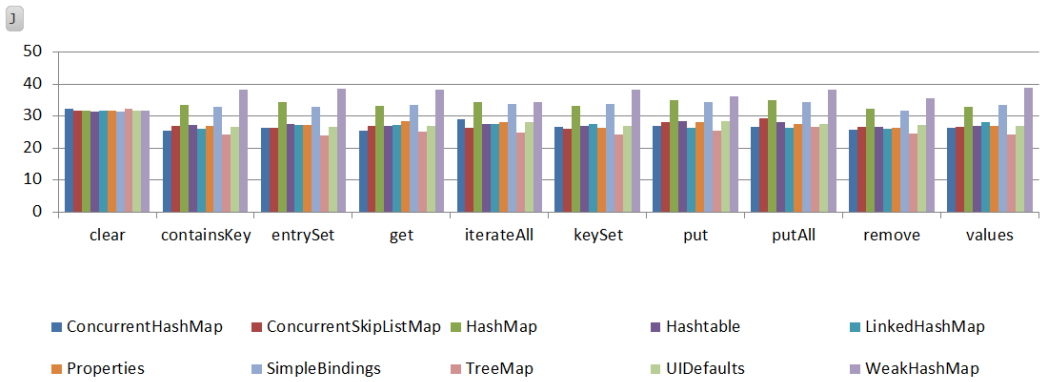

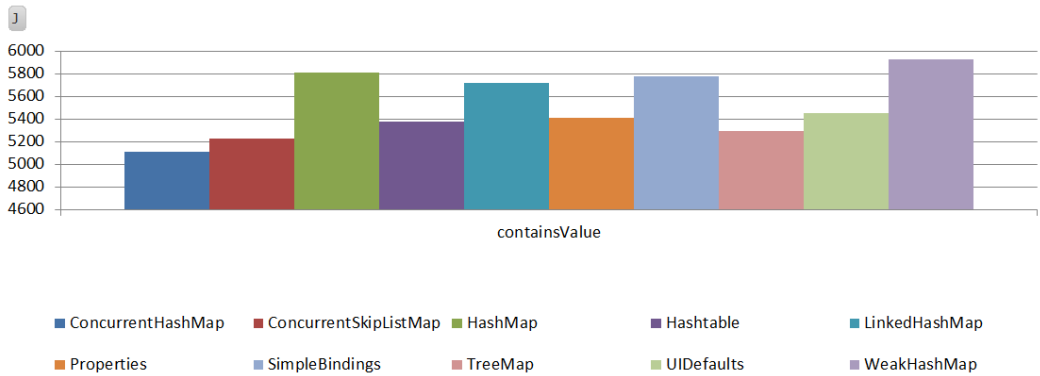

16



\end{document}